\documentclass[letterpaper, 10 pt, conference]{ieeeconf}  % Comment this line out if you need a4paper

\IEEEoverridecommandlockouts                              % This command is only needed if 
\usepackage{amsmath} % assumes amsmath package installed
\usepackage{amssymb}  % assumes amsmath package installed
\usepackage{caption}
\usepackage{subcaption}
\usepackage{graphicx}
\usepackage{algorithm}
\usepackage{algpseudocode}
\usepackage[utf8]{inputenc}
\usepackage{hyperref}
\usepackage{subcaption}
\usepackage{multirow}
\usepackage{multicol}
\usepackage{cite} % For citation commands
\usepackage{tikz}
\usepackage{tikz-3dplot}
\usepackage{subcaption}
\usepackage{tabularx}
\usepackage{booktabs}    % for cleaner table lines
\usepackage{array}       % for more column control
\providecommand{\keyword}[1]
{
  \small	
  \textbf{\textit{Keywords---}} #1
}

\title{\LARGE \bf
Learning based Modelling of Throttleable Engine Dynamics for Lunar Landing Mission
}

\author{Suraj Kumar$^{1}$$^{*}$, Aditya Rallapalli$^{1}$, Bharat Kumar GVP$^{1}$
\thanks{$^*$Corresponding Author}
\thanks{$^{1}$The authors are associated with Controls and Digital Area, U R Rao Satellite Center, Indian Space Research Organization, Bengaluru, Karnataka, India\{surajk, adityar, bharat\}@ursc.gov.in}
}

\begin{document}

\maketitle
\thispagestyle{empty}
\pagestyle{empty}

%%%%%%%%%%%%%%%%%%%%%%%%%%%%%%%%%%%%%%%%%%%%%%%%%%%%%%%%%%%%%%%%%%%%%%%%%%%%%%%%
\begin{abstract}

Typical lunar landing missions involve  multiple phases of braking to achieve soft-landing.  The propulsion system configuration for these missions consists of throttleable engines. This configuration involves complex interconnected hydraulic, mechanical, and pneumatic components each exhibiting non-linear dynamic characteristics. Accurate modelling of the propulsion dynamics is essential for analyzing closed-loop guidance and control schemes during descent. This paper presents a learning-based system identification approach for modelling of throttleable engine dynamics using data obtained from high-fidelity propulsion model. The developed model is validated with experimental results and used for closed-loop guidance and control simulations.

\end{abstract}
\keyword{\small System Identification, Engine dynamics, Supervised learning, Feature engineering}

%%%%%%%%%%%%%%%%%%%%%%%%%%%%%%%%%%%%%%%%%%%%%%%%%%%%%%%%%%%%%%%%%%%%%%%%%%%%%%%%
\section{Introduction}

Lunar landing missions require precise control of the spacecraft’s descent to ensure a safe and soft landing. These missions involve multiple braking phases and the propulsion system plays a critical role in velocity reduction and trajectory correction. Throttleable engines are commonly used for such missions due to their ability to modulate thrust levels based on guidance and control requirements. However, these propulsion systems comprise complex hydraulic, mechanical and pneumatic components, each exhibiting nonlinear dynamics and coupling effects. Accurate modelling of these dynamics is crucial for evaluating and refining closed-loop guidance and control strategies. 

Traditionally, high-fidelity propulsion system models are developed using fluid-mechanical equations and primarily employed for ground testing \cite{nath2017mathematical}. The application of machine learning techniques for system identification in aerospace systems has gained significant traction in recent years. Learning-based models provide an alternative by leveraging data-driven approaches to approximate system behavior, enabling its usage in simulation of closed-loop guidance and control strategies. In \cite{tong2020machine}, authors explore the integration of machine learning algorithms into the early stages of engine design, aiming to improve predictive capabilities and decision-making processes. In \cite{mcquarrie2021data}, authors presents a machine learning approach that combines data-driven learning with physics-based modeling to derive predictive reduced-order models for rocket engine combustion dynamics. The study in \cite{wright2024small}
demonstrates the application of machine learning to create a digital twin of a jet engine. Authors of \cite{peringal2024remaining} employs Long Short-Term Memory (LSTM) networks to predict the remaining useful life of jet engines. 

While traditional physics based models capture system behavior with high accuracy, their computational complexity makes them impractical for closed-loop guidance and control design, analysis and testing such as Six-Degrees-Of-Freedom (6DOF) and Onboard-in-loop (OILS) testing\cite{rallapalli2024landing}\cite{chaitanya2024chandrayaan}. To address this challenge, we propose a learning-based system identification approach to model throttleable engine dynamics. Using data from high-fidelity propulsion models developed for propulsion system testing, we develop a computationally efficient representation of engine behavior that can be integrated into closed-loop guidance and control analysis. The resultant system dynamics is represented as multi-input, multi-output (MIMO) finite-time difference model. In this framework, the current engine thrust, fuel consumption and gas pressure depends on the commanded thrust input and the history of measurable system states, such as mass flow rate and pressure. %In addition to modeling thrust output, the same approach is applied to develop dynamic models for chamber pressure and propellant mass consumption.

The remainder of the paper is organized as follows:
Sec. \ref{prop_sys_model} presents the details and configuration of propulsion system model considered for this study. Sec. \ref{input_excitation} presents the details of input excitation methodology used to generate dataset for training. Sec. \ref{feature_engineering} discusses feature engineering. Sec .\ref{model_formulation} presents the formulation of learning based model. Sec \ref{results} presents the verification and validation of model with standard inputs and descent profile. Sec \ref{conclusions} summarizes the major results of the paper .  
\section{Propulsion System Model}
\label{prop_sys_model}
\begin{figure}[h]
  \centering
  \includegraphics[width=0.5\textwidth]{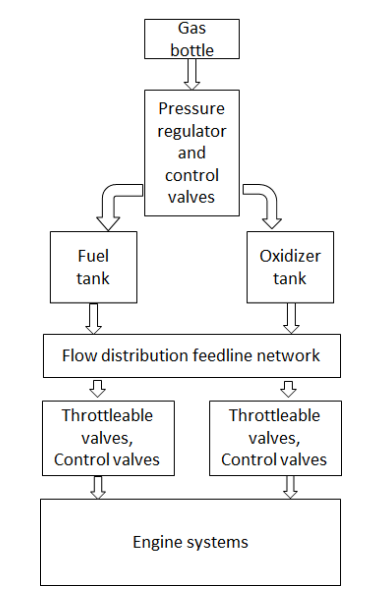}
  \caption{Propulsion System Schematics \cite{nath2017mathematical}}
  \label{fig:prop_sys}
\end{figure}

The Lunar Lander propulsion system considered here comprises of four throttleable engines of 800N rating each. The propulsion system is pressure fed through the fuel and oxidizer tanks that are being pressurized by a single Helium gas bottle. The system consists of three modules, namely gas module, fluid module and engine module. Gas module pressurizes the liquid module
propellant tanks that in turn produce flow to the engines
producing thrust. Fig \ref{fig:prop_sys} shows the functional schematics of propulsion system. \\
The gas module includes the gas bottle, check valves, latch valves, pressure regulators, and propellant tanks, with the pressure regulator modeled using a one-dimensional mechanical-fluid coupling approach. The fluid module begins at the propellant tanks and includes bends, feed lines, orifices, and valves. The engine module comprises the engine, solenoid valves, throttlable Flow Control Valves (TFCVs), and 800 N engines. Each component of the propulsion system is modeled in Simulink, resulting in a high-fidelity system dynamics. Detailed component level modelling details is available in \cite{nath2017mathematical}. It is impractical to incorporate this model for closed-loop simulation and testing due to computational load. This model is validated with experimental data and referred as high-fidelity propulsion model from which data-driven learning based model is derived for closed-loop simulation.

Our modeling approach begins with the design of input profiles that comprehensively span the dynamic operating range of throttleable engines. These inputs are then used to generate the corresponding outputs from the high-fidelity model that capture the system response under various conditions. The collected data is subsequently utilized to construct a learning-based model. To ensure reliability, the model is validated by comparing its predictions with the high-fidelity simulation results, followed by iterative refinement of model parameters. 
\section{Input Excitation and Data generation}
\label{input_excitation}
%Precise system identification requires a well-designed excitation signal that captures the full dynamic range of the system. A commonly used method is the sine sweep, which is primarily designed for frequency response analysis, such as obtaining the Bode response to determine the system’s bandwidth. However, in this study, our objective is different—we aim to mathematically represent the time-domain response of several dynamic variables, including mass flow rate, thrust, and pressure, rather than just the frequency characteristics of the system. While a conventional sine sweep covers a broad frequency spectrum, it typically employs a monotonic frequency increase at a fixed amplitude, leading to an extensive dataset that may not be optimal for training learning-based models. The sheer volume of data generated can make training computationally expensive without necessarily improving the model’s ability to capture the system's key dynamic behaviors. 
For a precise identification of the system, the excitation signal must capture the full dynamic range. Traditional sine sweeps, designed for frequency response analysis, focus on bandwidth estimation via Bode-like responses. However, our goal is to model the time-domain response of key variables such as mass flow rate, thrust, and pressure. Standard sine sweeps generate large datasets by monotonically increasing frequency at a fixed amplitude, which can be computationally expensive without significantly improving the information content for learning-based models. Instead, we adopt a more efficient excitation approach that reduces data volume while preserving essential system dynamics.

In this work, a four-dimensional excitation function corresponding to four engines is used to generate test inputs that span the entire operating range of the engine. This goal is to excite the system across different thrust levels and a range of frequencies for comprehensive system characterization.

The throttleable engine operates within a thrust range:

\begin{equation}
E_{\min} \leq E \leq E_{\max}
\end{equation}

To analyze system behavior, this range is discretized into M discrete thrust levels:

\begin{equation}
E_k = E_{\min} + \frac{k}{M} (E_{\max} - E_{\min}), \quad k = 0,1,2,\dots, M-1
\end{equation}

For each discretized thrust level \( E_k \), the system must be excited with a range of frequency variations to capture its full response. To excite the system across different operating conditions, the excitation function is defined as:

\begin{equation}
\begin{bmatrix} S_1 \\ S_2 \\ S_3 \\ S_4 \end{bmatrix} =
\begin{bmatrix} 
\cos(r) \cos(\theta) \cos(\phi) \\ 
\cos(r) \cos(\theta) \sin(\phi) \\ 
\cos(r) \sin(\theta) \\ 
\sin(r) 
\end{bmatrix}
\end{equation}

where:

\begin{align*}
r &= \pi t, \\
\theta &= \pi \sin(2 \sin(2 t)), \\
\phi &= \pi \sin(2 t).
\end{align*}
The inputs generated are scaled and biased to ensure that they span the operating range of the engine:

\begin{equation}
E_j (t, k) = E_{bias} + A_{\text{amp}} S_j(t), \quad j \in \{1,2,3,4\}
\end{equation}
where \( A_{\text{amp}} \) is the amplitude scaling factor;\( E_{bias} \) corresponds to a specific thrust level;\( S_j(t) \) represents the excitation signal component. Thus, for each thrust level \( E_k \), the system is excited with a range of frequency-modulated inputs. We also generate step and ramp inputs at various thrust levels and include that in the input space of excitation. This approach ensures that the learned model captures both thrust-dependent variations and frequency-dependent dynamics.
\section{Feature Engineering}
\label{feature_engineering}
The selection of the state of the system for the input features is guided by domain knowledge and expert analysis. Rather than relying on deep neural networks to automatically discover features, we leverage the underlying physics of the problem to design input features. This turns out to be more data-efficient learning approach. These features comprise both current and history of states as follows: 
\begin{itemize}
    \item \(\ T_r, T_{r-h} \) – Reference thrust input and its history.
    \item \(\ T_{o-h} \) – Output thrust history.
    \item \(\ P, P_h \) – gas pressure and its history.
    %\item \(\ \dot{m}_{f-h} \) – History of fuel mass flow rate.
    %\item \(\ \dot{m}_o^h \) – History of oxidizer mass flow rate.
    \item \(\ m_{f-h} \) – History of ejected fuel mass.
    \item \(\ m_{o-h} \) – History of ejected oxidizer mass.
    \item \(\ S_e \) – Engine On-Off status.
    \item \(\ \lambda = (m_f + m_o)^{-1} \) – Scaled inverse of total ejected mass.
\end{itemize}

\section{Learning based model Formulation}
\label{model_formulation}
To capture the temporal dependencies of states in dynamics, we define an extended state representation for each individual feature. Given a time history of length \( n \), the extended state representing the history of generic feature \( x^h \) at time step \( t \) is defined as:

\begin{equation}
    {x}_h(t) = \begin{bmatrix} x_{t-1} & x_{t-2} & \dots & x_{t-n} \end{bmatrix}^T.
\end{equation}
Here $n$ denotes the history length for the state $x$. While the history length can vary for each state, a uniform history length is adopted for all states to maintain the tractability of the learning problem.
%Using this notation, we construct the extended state vectors for all %features as:

%\begin{align*}
%    \tilde{T}_r &= \begin{bmatrix} T_r(t) & T_r(t-1) & \dots & T_r(t-n) \end{bmatrix}^T, \\
%    \tilde{T}_h &= \begin{bmatrix} T_h(t) & T_h(t-1) & \dots & T_h(t-n) \end{bmatrix}^T, \\
%    \tilde{P}_h &= \begin{bmatrix} P_h(t) & P_h(t-1) & \dots & P_h(t-n) \end{bmatrix}^T, \\
 %   \tilde{m}_f &= \begin{bmatrix}  m_f(t-1) & \dots & m_f(t-n) \end{bmatrix}^T, \\
 %   \tilde{m}_o &= \begin{bmatrix}  m_o(t-1) & \dots & m_o(t-n) \end{bmatrix}^T, \\
%    \tilde{\dot{m}}_f &= \begin{bmatrix}  \dot{m}_f(t-1) & \dots & \dot{m}_f(t-n) \end{bmatrix}^T, \\
%    \tilde{\dot{m}}_o &= \begin{bmatrix} \dot{m}_o(t-1) & \dots & \dot{m}_o(t-n) \end{bmatrix}^T, \\
%    \tilde{\Lambda} &= \begin{bmatrix} {\lambda}(t-1) & \dots & {\lambda}(t-n) \end{bmatrix}^T.
%\end{align*}

The engine On-Off status \( S_e \) is included without time history, since it is a discrete switching variable:

\begin{equation*}
    S_e(t) = \begin{bmatrix} S_e(t) \end{bmatrix}.
\end{equation*}

Using these extended states, the input feature vector denoted by $\tilde{\mathbf{X}}_t$, used for training is given as:

\begin{equation*}
    \tilde{\mathbf{X}}_t = \begin{bmatrix} T_r & T_{r-h}^T & T_{o-h}^T & P & P_h^T & m_{f-h}^T & m_{o-h}^T & S_e & \lambda \end{bmatrix}^T.
\end{equation*}
The output vector, denoted as $\tilde{\mathbf{Y}}_t$, represents the predicted output from the model and is defined as:
\begin{equation*}
    \tilde{\mathbf{Y}}_t = \begin{bmatrix} T_o^T & P & m_f^T & m_o^T \end{bmatrix}^T.
\end{equation*}
where $T_o^T \in \mathcal{R}^4$ represent the output thrust for each engines; $m_o,m_f$ represent the ejected oxidizer and fuel mass; $P$ represents the gas pressure. 

The collected input-output data obtained from the high-fidelity model is then appropriately transformed into this format. At each time step
$t$, the input vector $\tilde{\mathbf{X}}_t$ is constructed incorporating both the current and the history of states states. Similarly, the corresponding output vector $\tilde{\mathbf{Y}}_t$ is extracted, forming training set, $\mathcal{D}$, given as:

\begin{equation*}
\label{tgo_set}
\mathcal{D} = \{(\tilde{X}_i, \tilde{Y}_i) \mid \tilde{X}_i \in \mathcal{R}^{11n+10}, \tilde{Y}_i \in \mathcal{R}^{7}, i =1,2,.., |\mathcal{D}|\}
\end{equation*}
The relationship between input and output vector is expressed as:
\begin{equation}
     \tilde{\mathbf{Y}}_t = \mathbf{K} \phi(\tilde{\mathbf{X}}_t; \mathcal{D})
\end{equation}
where $\mathbf{K} \in \mathcal{R}^{7 \times 11n+10}$ represent the coefficient matrix to be learned from the data; $\phi$ represent the polynomial function approximator. 

The coefficient matrix K is arrived at using lasso regression on dataset $\mathcal{D}$ as
\begin{equation*}
\min_{\mathbf{K}} \left( \frac{1}{2} \sum_{i=1}^{|\mathcal{D}|} \left( \tilde{\mathbf{Y}}_t(i) - K \phi(\tilde{\mathbf{X}}_t(i)) \right)^2 + \mu \|K\|_1 \right)
\end{equation*}
The first term in the loss function represents the reconstruction error between input and output. The second term represents the $l_1$-norm regularization that promotes sparsity in the solution.  
$\mu \|K\|_1$  is the regularization term controlling the strength of the sparsity. Specifically, $l_1$-norm regularization promotes sparsity in the solution as it gives relatively large weight to small residuals, and therefore results in many optimal residuals small, or even zero\cite{boyd2004convex}. Therefore, it implicitly select appropriate features that best captures the sensitive parameters in the dataset and also prevent overfitting.

The appropriate choice of critical hyper-parameters such as order history length (n) and regularization term ($\mu$) significantly affects the performance of learning model. Optimal choice for these parameters is arrived by carrying out parametric analysis. In order to evaluate the performance during parametric analysis, the available dataset is divided into training and testing subsets and k-fold cross-validation approach is employed to compute Root Mean Square Error(RMSE) between the trained and true model. 
 
\begin{figure}[h]
  \centering
  \includegraphics[width=0.45\textwidth]{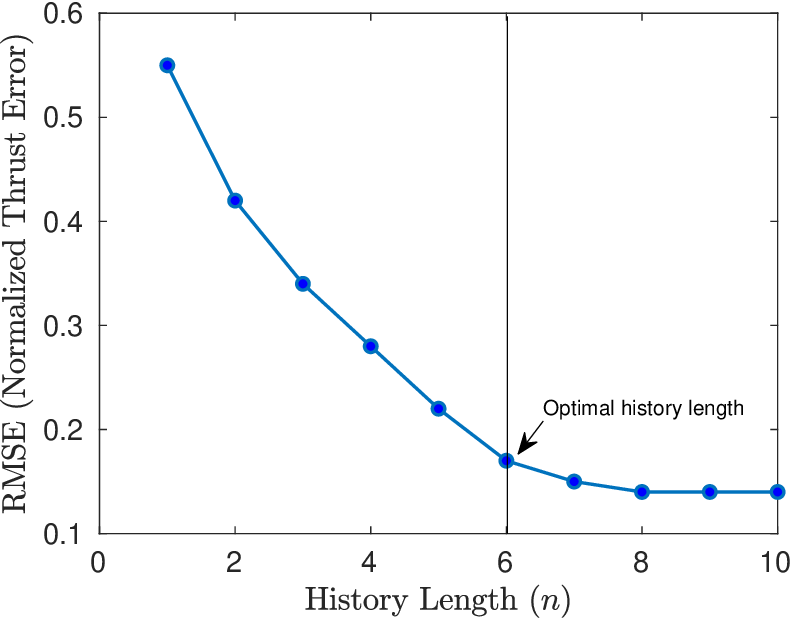}
  \caption{Model Error as function of history length}
  \label{fig:rmse_n}
\end{figure}

\begin{figure}[h]
  \centering
  \includegraphics[width=0.45\textwidth]{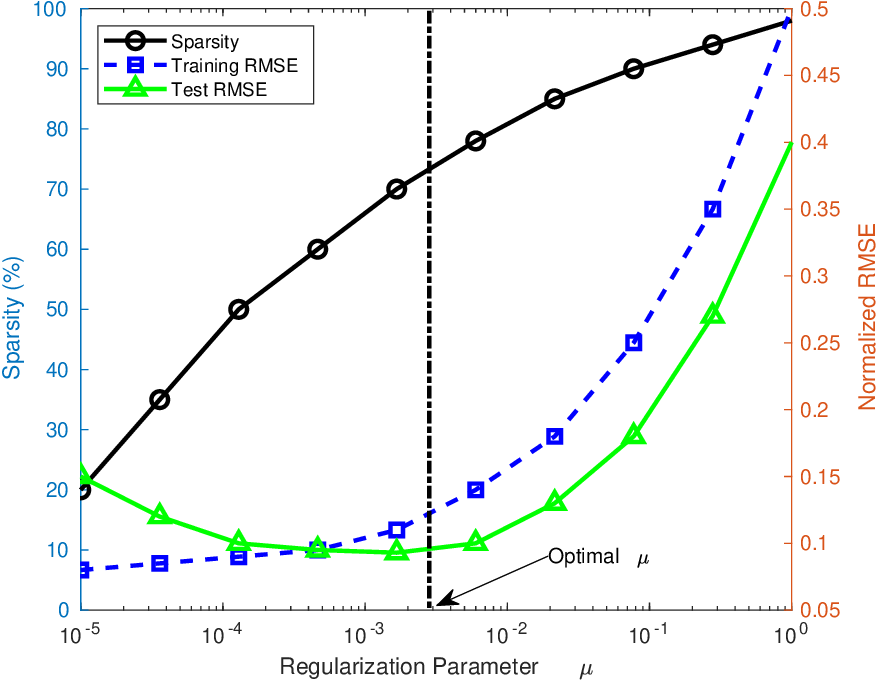}
  \caption{Model Error as function of sparsity}
  \label{fig:spar}
\end{figure}

Fig \ref{fig:rmse_n} shows the model error as history length is varied parametrically. History length of 6 is selected as it results in minimum error between trained and true model. Similarly, the regularization parameter $\mu$ is varied logarithmically from $10^{-5}$ to $10^{0}$ to analyze the impact on model sparsity and prediction error. As shown in Fig \ref{fig:spar}, increasing $\mu$ from $10^{-5}$ to $10^{-2}$ resulted in minimal accuracy loss ($<1.5\%$) while increasing model sparsity from $20\%$ to $78\%$. However, beyond $\mu \approx 10^{-2}$, the test RMSE increased significantly, indicating that excessive feature elimination led to underfitting. The training RMSE, in contrast, remained low for small $\mu$, highlighting overfitting at very low regularization values. This trade-off follows a Pareto-like behavior, where an optimal balance is achieved at $\mu_1 \approx 10^{-2}$.
\begin{figure}[h]
  \centering
  \includegraphics[width=0.5\textwidth]{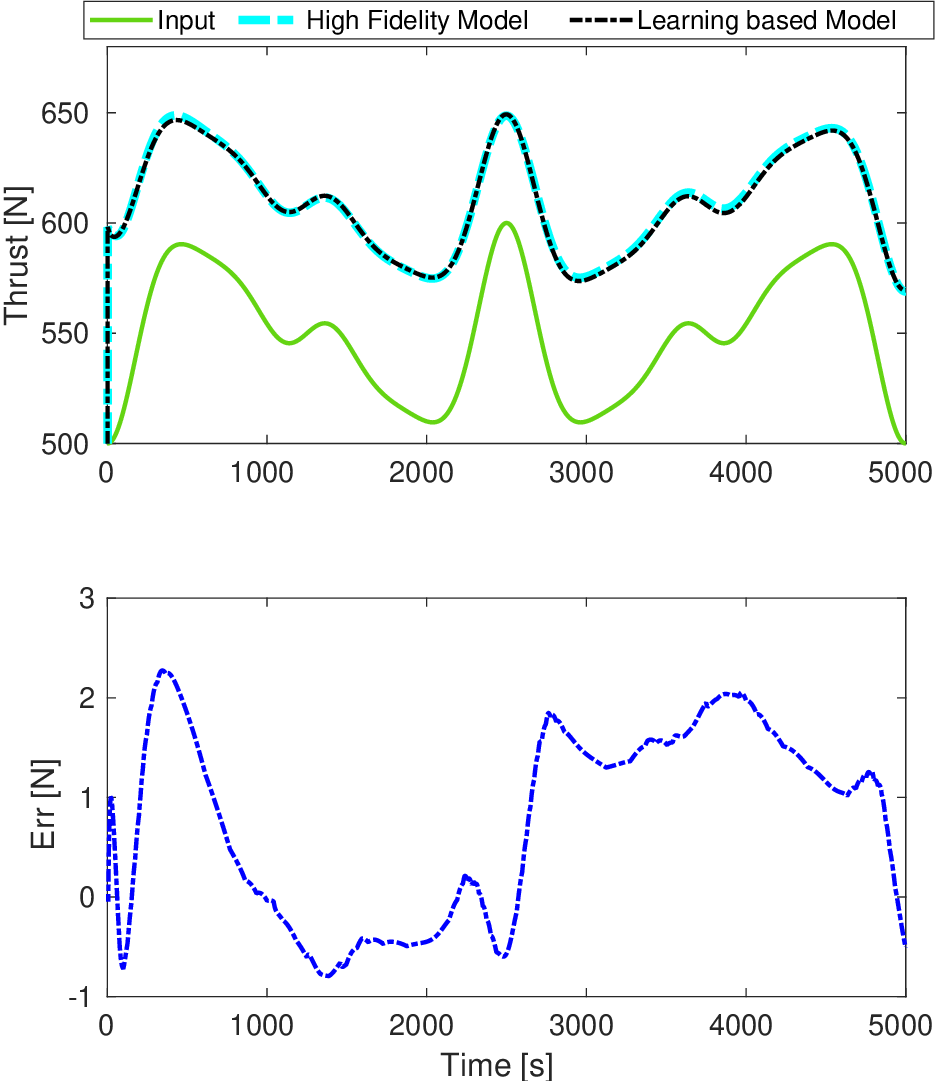}
  \caption{Sinusoidal sweep response}
  \label{fig:sine_response}
\end{figure}
\section{Results}
\label{results}
The proposed learning-based model for throttleable engine dynamics is evaluated on test signals by comparing the step and sine response against the high-fidelity model and finally with the descent trajectory profile. 

Sinusoidal signals at amplitudes 400-700 N and several combinations of such sinusoidal sweep is obtained from projecting 4D sphere of radius 400-700 N into 3D vector space and results for learning based model output for such inputs are compared with high fidelity model output. Fig \ref{fig:sine_response} shows the comparison between models for one particular case where sinusoid signals at amplitude 600N applied to both models. It is observed that learning based model captures the true dynamics within error bound of $\pm2N$. \\

\begin{figure}[h]
  \centering
  \includegraphics[width=0.5\textwidth]{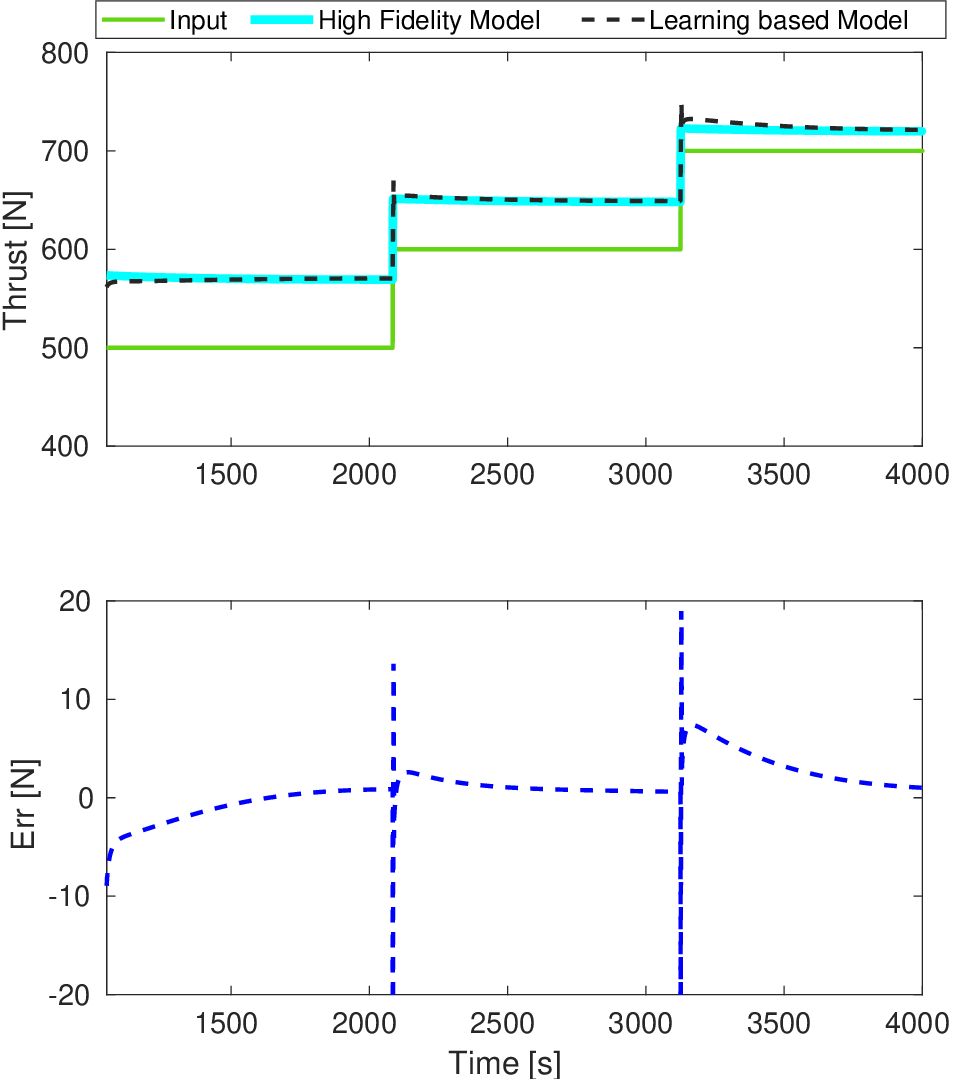}
  \caption{Step-stair response}
  \label{fig:rise_response}
\end{figure}

\begin{figure}[h!]
  \centering
  \includegraphics[width=0.5\textwidth]{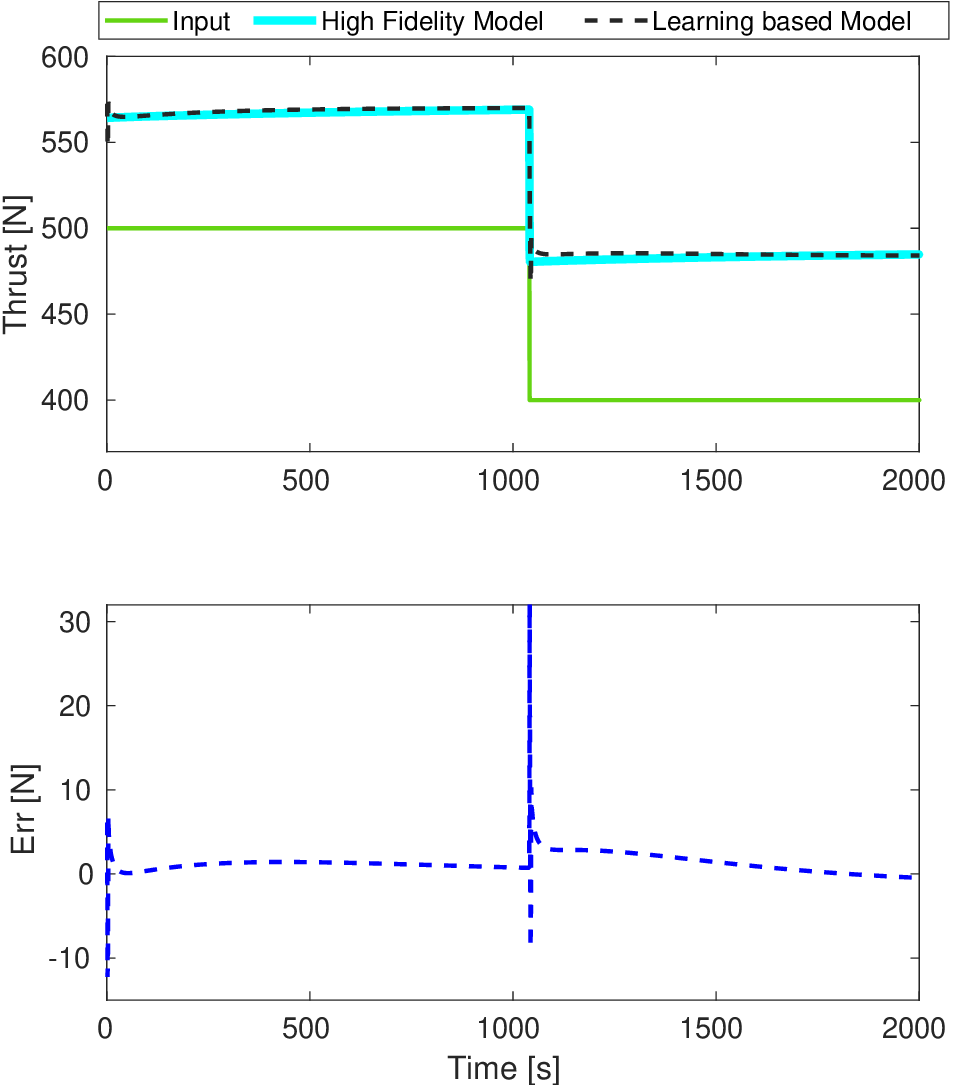}
  \caption{Fall response}
  \label{fig:fall_response}
\end{figure}

\begin{figure}[h]
  \centering
  \includegraphics[width=0.5\textwidth]{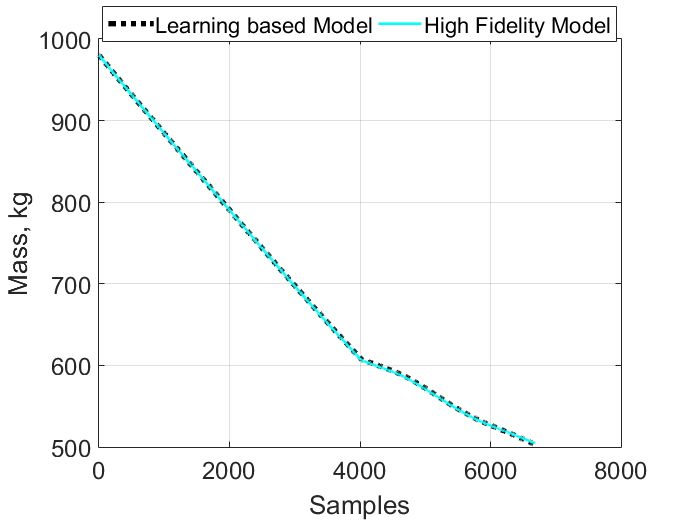}
  \caption{Total Module Mass}
  \label{fig:mass}
\end{figure}

A step-stair input signal with varying amplitudes is applied to analyze the response of the model and compare its predicted outputs against the true behavior of the system, as shown in Fig \ref{fig:rise_response}. The model error is observed to be bounded within $\pm20N$ in transient and within $\pm2N$ in steady state. For closed-loop simulation, transient error is injected as uncertainty in the model to perform Monte Carlo simulation. \\
Finally, fall response of the model is evaluated against true model as shown in Fig \ref{fig:fall_response}. The difference in behavior between fall and rise response is evident from the fact that the overall system is non-linear. \\
\begin{figure}[h!]
  \centering
  \includegraphics[width=0.5\textwidth]{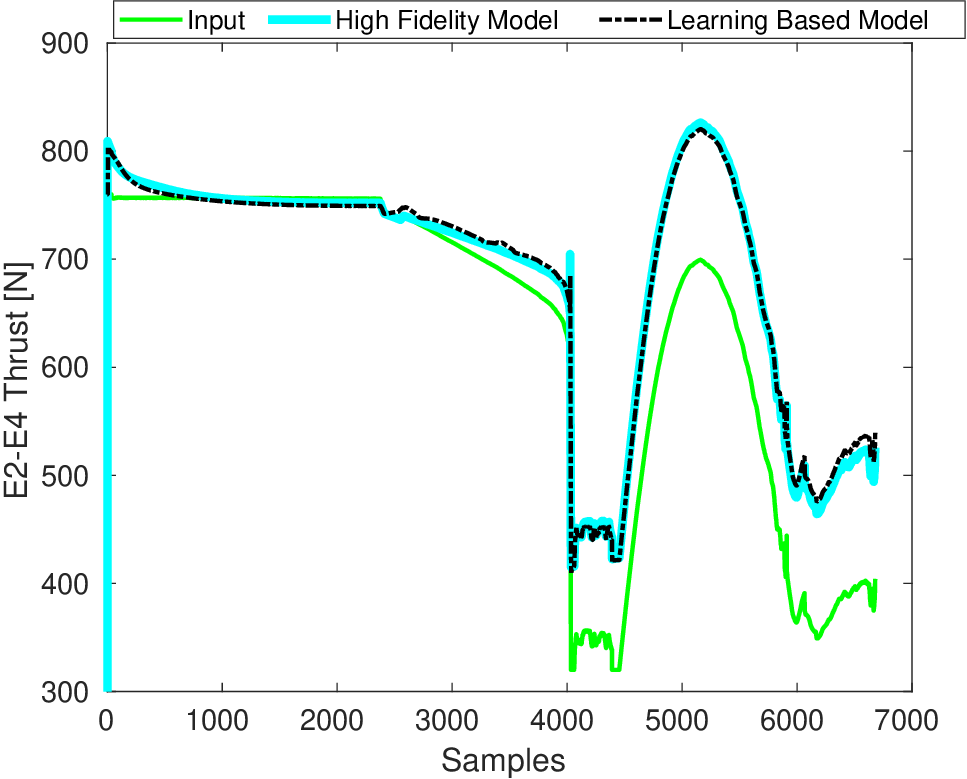}
  \caption{Engine 2 and 4 Thrust}
  \label{fig:E1-E3}
\end{figure}
 \begin{figure}[h!]
  \centering
  \includegraphics[width=0.5\textwidth]{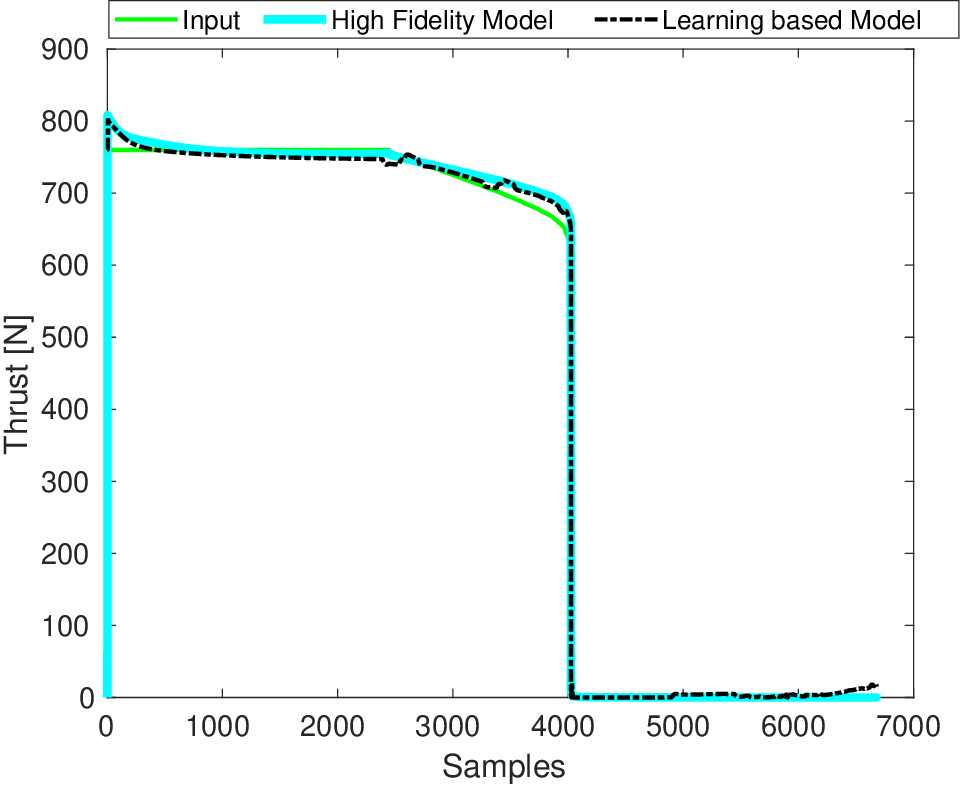}
  \caption{Engine 1 and 3 Thrust}
  \label{fig:E2-E4}
\end{figure}
After the model was validated against the high-fidelity model and refined its coefficients, the model was further evaluated by using an open-loop thrust profile for the entire powered descent trajectory. Fig \ref{fig:E1-E3}-\ref{fig:E2-E4} shows the outputs obtained for 4 engines from the learning-based model and the high-fidelity model. It is evident that our model closely approximates the true system dynamics within $\pm15N$ model error. For closed-loop simulation model error is fed as model uncertainty to carry out monte-carlo simulation. Finally, Fig \ref{fig:mass} shows the total module mass computed from the oxidizer and fuel mass for the entire powered descent trajectory. Mass computation is highly accurate as compared to high fidelity model with maximum error within $\pm1.5 kg$.
 
\pagebreak

\section{Conclusions}
\label{conclusions}
The proposed learning-based model has proven to be effective in predicting both thrust output and fuel consumption compared to the high-fidelity propulsion system model. The model successfully captures the non-linear behavior of the engine, with close agreement between predicted and actual values across different input cases. The model closely approximates the 'true' model for various mission profiles obtained with closed-loop simulation with a variation of around $15-20N$. The results also demonstrate that the model performs well on unseen inputs, indicating a strong generalization capability. However, the transient error in the step response suggests that a further re-tuning of the features or regularization settings could improve the accuracy. In the future, our aim is to improve the accuracy of transient error by further incorporating more expressive models such as deep neural networks. 

\section*{Acknowledgments}
We convey our sincere gratitude to the Indian Space Research Organization (ISRO) for encouraging and supporting this research. We wish to gratefully acknowledge the excellent review committee at ISRO for reviewing this work and providing valuable feedback. We acknowledge the support of the Propulsion Research and Studies Entity, LPSC, Valiamala, for providing the Simulink model of the propulsion system and for their contributions to the development of feature engineering for training. This research work was carried out at the U.R. Rao Satellite Centre, ISRO as a part of GNC development for the Chandrayaan-2 mission.

% Generated by IEEEtran.bst, version: 1.14 (2015/08/26)

\end{document}